\documentstyle[aps,graphicx]{revtex}
\begin{document}
\title{$Y$-Scaling Analysis of the Deuteron Within the
Light-Front Dynamics Method}
\author{M.K. Gaidarov, M.V. Ivanov, and A.N. Antonov}
\address{Institute of Nuclear Research and Nuclear Energy,\\
Bulgarian Academy of Sciences, Sofia 1784, Bulgaria}
\maketitle

\begin{abstract}
The concept of relativistic scaling is applied to describe the
most recent data from inclusive electron-deuteron scattering at
large momentum transfer. We calculate the asymptotic scaling
function $f(y)$ of the deuteron using its relationship with the
nucleon momentum distribution. The latter is obtained in the
framework of the relativistic light-front dynamics (LFD) method,
in which the deuteron is described by six invariant functions
$f_{i}$ ($i$=1,...,6) instead of two ($S$ and $D$ waves) in the
nonrelativistic case. Comparison of the LFD asymptotic scaling
function with other calculations using $S$ and $D$ waves
corresponding to various nucleon-nucleon potentials, as well as
with the Bethe-Salpeter result is made. It is shown that for
$|y|>$ 400 MeV/c the differences between the LFD and the
nonrelativistic scaling functions become larger.
\end {abstract}
\vspace{1cm}

\section{Introduction}
High-energy electron scattering from nuclei can provide important
information on the wave function of nucleons in the nucleus. In
particular, using simple assumptions about the reaction mechanism,
scaling functions can be deduced that, if shown to scale (i.e., if
they are independent of the momentum transfer), can provide
information about the momentum and energy distribution of the
nucleons. Several theoretical studies
\cite{Ji89,Cio91,Cio94,Cio96,Gai95} have indicated that such
measurements may provide direct access to studies of short-range
nucleon-nucleon (NN) correlation effects.

Since West's pioneer work \cite{West75}, there has been a growth
of interest in $y$-scaling analysis, both in its experimental and
theoretical aspects. This is motivated by the importance of
extracting nucleon momentum distributions from the experimental
data. West showed that in the impulse approximation, if
quasielastic scattering from a nucleon in the nucleus is a
dominant reaction mechanism, a scaling function $F(y)$ could be
extracted from the measured cross section which is related to the
momentum distribution of the nucleons. In the simplest
approximation the corresponding scaling variable $y$ is the
minimum momentum of the struck nucleon along the direction of the
virtual photon. In principle, the scaling function $F(y,Q^{2})$
depends on both $y$ and momentum transfer $Q^{2}$ (-$Q^{2}$ is the
square of the four-momentum transfer), but at sufficiently high
$Q^{2}$ values the dependence on $Q^{2}$ should vanish yielding
scaling. However, any attempt to extract the momentum distribution
from the $y$-scaling in electron-nucleus scattering faces the
problem of estimating both effects from the final-state
interactions (FSI) of the struck nucleon with the rest of the
nucleus and from the nucleon binding. Previous calculations
\cite{Ben91,Ben93,Cio2001} suggest that the contribution from the
final-state interactions should vanish at sufficiently high
$Q^{2}$. The FSI lead to sizable scaling violation effects only at
low values of the three-momentum transfer $|{\bf
q}|=\sqrt{Q^{2}+\nu^{2}}$ ($\nu$ is the energy loss)
\cite{Cio2001,Cio87}. The most important dynamical effects, such
as binding corrections, which represent the fact that for complex
nuclei the final spectator $A-1$ system can be left in all
possible excited states including the continuum, have been treated
in \cite{Cio91} in terms of spectral functions. This problem has
been solved in \cite{Cio99LA,Far99} by introducing a new scaling
variable which gives direct, global and independent of $A$ link
between the experimental data and the longitudinal momentum
components.

Recently inclusive electron scattering has been studied at the
Thomas Jefferson National Accelerator Facility (TJNAF) with 4.045
GeV incident beam energy from C, Fe and Au targets \cite{Arr99} to
$Q^{2}$ $\approx$ 7 (GeV/c)$^{2}$. Data were also taken using
liquid targets of hydrogen and deuterium \cite{Arr2002}. The data
presented in \cite{Arr99,Arr2002} represent a significant increase
of the $Q^{2}$ range compared to previous SLAC measurement
\cite{Day87}, in which an approach to the scaling limit for heavy
nuclei is suggested but for low values of $|y|<$ 0.3 GeV at
momentum transfers up to 3 (GeV/c)$^{2}$, and, moreover, a scaling
behaviour is observed for the first time at very large negative
$y$ ($y$= -0.5 GeV/c). From theoretical point of view the extended
region of $y$ measured at TJNAF is of significant importance since
this is a regime where the nucleon momentum distribution is
expected to be dominated by short-range NN correlations. On the
other hand, it is interesting to note that contributions from
short-range FSI may also result in a scaling-like behavior due to
the small $Q^{2}$ dependence of these effects, and that these
contributions are also dominated by short-range correlations.
Obviously, a complete understanding of this electron-nucleus
scattering requires a relativistic approach to the quantities
related to the $y$-scaling analysis for a detailed comparison with
the experimental data.

A relativistic $y$-scaling has been considered in \cite{Cio99} by
generalizing the nonrelativistic scaling function to the
relativistic case. Realistic solutions of the spinor-spinor
Bethe-Salpeter (BS) equation for the deuteron with realistic
interaction kernel were used for systematic investigation of the
relativistic effects in inclusive quasielastic electron-deuteron
scattering. The approach of $y$-scaling presented in \cite{Cio99}
is fully covariant, with the deuteron being described by eight
components, namely the $^{3}S_{1}^{++}$, $^{3}D_{1}^{++}$,
$^{3}S_{1}^{--}$, $^{3}D_{1}^{--}$, $^{1}P_{1}^{+-}$,
$^{1}P_{1}^{-+}$, $^{3}P_{1}^{+-}$, and $^{3}P_{1}^{-+}$ waves.
The first two waves directly correspond to the $S$ and $D$ waves
in the deuteron, with the waves with negative energy vanishing in
the nonrelativistic limit. It has been demonstrated in
\cite{Cio99} that, if the effects from the negative energy
$P$-states are disregarded, the concept of covariant momentum
distribution can be introduced.

Recently a successful relativistic description of the nucleon
momentum distribution in deuteron has been done \cite{Ant2002}
within the light-front dynamics method \cite{Car95,Car98}. The
most important result from the calculations in \cite{Ant2002} is
the possibility of the LFD method to describe simultaneously both
deuteron charge form factors (that has been shown in \cite{Abo1})
and the momentum distribution. It is shown in \cite{Bon99,Car98}
that after the projection of the Bethe-Salpeter amplitude on the
light front, the six components of the LFD deuteron wave function
are expressed through integrals over the eight components of the
deuteron Bethe-Salpeter amplitude. Provided the nucleon-nucleon
interaction is the same, these approaches incorporate by different
methods the same relativistic dynamics. The wave functions in LFD
are the direct relativistic generalization of the nonrelativistic
ones in the sense that they are still the probability amplitudes.
Therefore they can be used in the relativistic nuclear physics
(e.g. \cite{Car95}).

The aim of our work is using the nucleon momentum distribution
$n(k)$ obtained with the LFD method to calculate the deuteron
scaling function. The result for the asymptotic function is
compared with the recent TJNAF data measured at six values of
$Q^{2}$. In particular, the scaling behavior observed for very
large negative $y$ providing momenta higher than those
corresponding to existing experimental data for $n(k)$ may allow
to distinguish the properties of the covariant LFD method from the
potential approaches. The comparison with the BS result for the
scaling function serves as a test for the consistency of both
covariant approaches treating the deuteron relativistically in the
case of $y$-scaling.

The paper is organized as follows. In Section II the definition
and the physical meaning of the scaling variable and the scaling
function are briefly reviewed together with some basic relations
between the nucleon spectral function, the scaling function and
the momentum distribution. The results for the nucleon momentum
distribution in deuteron obtained within the LFD method are given
in Section III. The calculated LFD asymptotic scaling function of
the deuteron is presented in Section IV, where a comparison with
the Bethe-Salpeter result and with some nonrelativistic
calculations is also done. The summary of the present work is
given in Section V.

\section{Basic relations in the $y$-scaling method}

The scaling function is defined as the ratio of the measured cross
section to the off-shell electron-nucleon cross section multiplied
by a kinematic factor:
\begin{equation}\label{eq:scf}
\nonumber F(q,y)=\frac{d^{2}\sigma}{d\Omega d\nu}
(Z\sigma_{p}+N\sigma_{n})^{-1}\frac{q}{[M^{2}+(y+q)^{2}]^{1/2}},
\end{equation}
where $Z$ and $N$ are the number of protons and neutrons in the
target nucleus, respectively, $\sigma_{p}$ and $\sigma_{n}$ are
the off-shell cross sections, and $M$ is the mass of the proton.
In analysing quasielastic scattering in terms of the $y$-scaling a
new variable $y=y(q,\nu)$ is introduced. Then the nuclear
structure function which is determined using the spectral function
$P(k,E)$ as
\begin{equation}
F(q,\nu)=2\pi\int_{E_{min}}^{E_{max}(q,\nu)}dE
\int_{k_{min}(q,\nu,E)}^{k_{max}(q,\nu,E)}kdkP(k,E),
\label{eq:strf}
\end{equation}
can be expressed in terms of $q$ and $y$ rather than $q$ and $\nu$
(see Eq. (\ref{eq:scf})). In Eq. (\ref{eq:strf})
$E=E_{min}+E^{*}_{A-1}$ is the nucleon removal energy with
$E^{*}_{A-1}$ being the excitation energy of the final $A-1$
nucleon system.

The most commonly used scaling variable $y$ is obtained
\cite{Pace82} starting from relativistic energy conservation,
setting $k=y$, $\frac{{\bf k}{\bf q}}{kq}=1$ and the excitation
energy $E^{*}_{A-1}$=0, and is defined through the equation
\begin{equation}
\nu+M_{A}=(M^2+q^2+y^2+2yq)^{1/2}+(M_{A-1}^2+y^2)^{1/2},
\label{eq:yvar}
\end{equation}
where $M_{A}$ is the mass of the target nucleus and $M_{A-1}$ is
the mass of the $A-1$ nucleus. Therefore, $y$ represents the
longitudinal momentum of a nucleon having the minimum removal
energy ($E=E_{min}$, i.e.  $E^{*}_{A-1}$=0).

At high values of $q$ a pure scaling regime is achieved, where
$k_{min}\approx|y-(E-E_{min})|$ and Eq. (\ref{eq:strf}) becomes
\begin{equation}
F(q,y)\rightarrow f(y)=2\pi\int_{E_{min}}^{\infty}dE
\int_{|y-(E-E_{min})|}^{\infty}kdkP(k,E).
\label{eq:limit}
\end{equation}
In Eq. (\ref{eq:limit}) the particular behavior of $P(k,E)$ at
large $k$ and $E$ is used in order to extend the upper limits of
integration to infinity \cite{Cio91}.

In the deuteron one always has $E^{*}_{A-1}$=0, so that the
spectral function is entirely determined by the nucleon momentum
distribution $n(k)$, i.e. $P(k,E)=n(k)\delta(E-E_{min})$, and,
consequently, $k_{min}=|y|$ for any value of $q$. The scaling
function (\ref{eq:limit}) reduces to the longitudinal momentum
distribution
\begin{equation}
f(y)=2\pi\int_{|y|}^{\infty}k\;dk\;n(k).
\label{eq:long}
\end{equation}

\section{Nucleon momentum distribution in the light-front dynamics method}

The relativistic deuteron wave function (WF) on the light-front
$\Psi(\vec{k},\vec{n})$ depends on two vector variables: i) the
relative momentum $\vec{k}$ and ii) the unit vector $\vec{n}$
along $\vec{\omega}$ which determines the position of the
light-front surface. Due to this, the WF is determined by six
invariant functions $f_{i}$ ($i$=1,...,6) instead of two ($S$- and
$D$-waves) in the nonrelativistic case. Each one of these
functions depends on two scalar variables $k$ and
$z=cos(\widehat{\vec{k},\vec{n}})$. In LFD these six functions are
calculated within the relativistic one-boson-exchange model. As
shown in \cite{Car95}, in the nonrelativistic limit the functions
$f_{3-6}$ become negligible, $f_{1,2}$ do not depend on $z$ and
turn into $S$- and $D$-waves (${f_{1}}{\approx}{u_{S}}$,
$f_{2}{\approx}-u_{D}$) and the wave function
$\Psi(\vec{k},\vec{n})$ becomes the usual nonrelativistic wave
function. One of the most important properties of the functions
$f_{1-6}$ found in \cite{Car95} is that for $k\geq$ 2$\div$2.5
fm$^{-1}$ the component $f_{5}$ (being related mainly to
$\pi$-exchange) exceeds sufficiently the $S$- and $D$-waves. This
fact is very important in the calculations of $n(k)$ in deuteron
as it will be shown below.

\begin{figure}[h]
\begin{center}
\includegraphics[width=80mm]{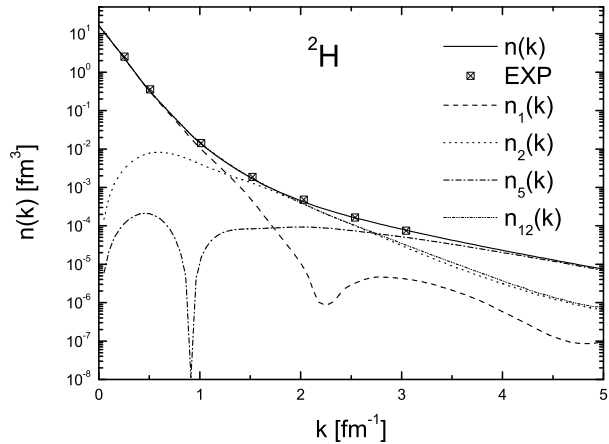}
\vspace{2mm}\caption{The nucleon momentum distribution in
deuteron. The contributions of $n_{1}$, $n_{2}$,
$n_{12}=n_{1}+n_{2}$ and $n_{5}$ are presented. The $y$-scaling
data are from \protect\cite{Cio91}. The normalization is:
$\int{{n(k)}d^{3}\vec{k}}=1$.}
\label{fig1}
\end{center}
\end{figure}

The LFD calculations have shown (for more details, see Ref.
\cite{Ant2002}) that, as expected, the most important
contributions to the total $n(k)$ give terms related to the
$f_{1}$, $f_{2}$ and $f_{5}$ functions
\begin{equation}\label{n1}
n(k){\simeq}n_{1}(k)+n_{2}(k)+n_{5}(k).
\end{equation}
The contributions of $n_{1}$, $n_{2}$, $n_{12}=n_{1}+n_{2}$ and
$n_{5}$ are compared in Fig. \ref{fig1}. It can be seen that,
while the functions $f_{1}$ and $f_{2}$ give a good description of
the $y$-scaling data of $n(k)$ for $k<2$ fm$^{-1}$ (like the $S$-
and $D$-wave functions in the nonrelativistic case), it is
impossible to explain the high-momentum components of $n(k)$ at
$k>2$ fm$^{-1}$ without the contribution of the function $f_{5}$.
We note that the deviation of the total $n(k)$ from the sum
$n_{12}=n_{1}+n_{2}$ starts at $k$ around 1.8 fm$^{-1}$. All this
shows the important role of NN interactions which incorporate
exchange of relativistic mesons in the case of the deuteron.

\begin{figure}[h]
\begin{center}
\includegraphics[width=80mm]{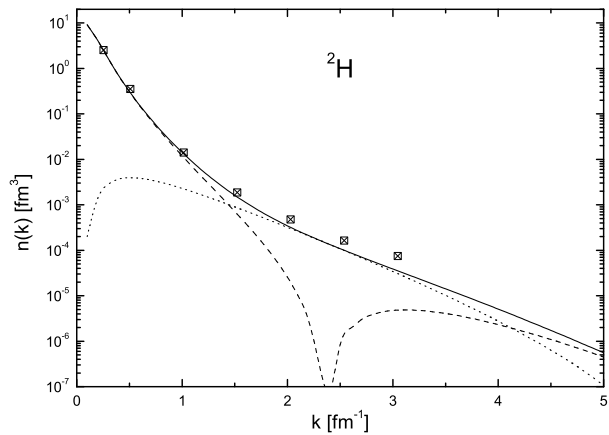}
\vspace{2mm}\caption{The nucleon momentum distribution in deuteron
(solid line) calculated using Eqs. (\ref{n10}) and (\ref{n11})
with $S$- and $D$-wave functions corresponding to the
charge-dependent Bonn potential \protect\cite{Mac01}. $S$- and
$D$-contributions are given by dashed and dotted line,
respectively. The $y$-scaling data are from \protect\cite{Cio91}.}
\label{fig2}
\end{center}
\end{figure}

The nucleon momentum distribution in the deuteron using $S$- and
$D$-wave functions $\Psi_{S}(k)$ and $\Psi_{D}(k)$ corresponding
to various NN potentials, such as the charge-dependent Bonn
potential \cite{Mac01}, the Argonne v$_{18}$ \cite{Wir95}, the
Nijmegen - I ,- II and - Reid 93 \cite{Sto94} and Paris 1980
\cite{Lac80} is also computed in \cite{Ant2002} by the expression:
\begin{equation}\label{n10}
n(k)=\frac{1}{4\pi}[\Psi_{S}^{2}(k)+\Psi_{D}^{2}(k)]\equiv
n_{S}(k)+n_{D}(k)
\end{equation}
with
\begin{equation}\label{n11}
\int{{n(k)}d^{3}\vec{k}}=1\,.
\end{equation}
In Fig. \ref{fig2} the result for $n(k)$ using the
charge-dependent Bonn potential \cite{Mac01} is given and compared
with the $y$-scaling data. As can be seen, the $D$-component of
$n(k)$ is important but even its inclusion does not give a very
good agreement with the data for $k\geq 2$ fm$^{-1}$. In the next
Fig. \ref{fig3} we present the LFD result for $n(k)$ compared with
the calculations using the WF's corresponding to Nijmegen-I,-II,
-Reid 93, Argonne v$_{18}$ and Paris 1980 NN potentials and with
the $y$-scaling data. We would like to note that: i) the results
of the calculations using the NN potentials, such as Nijmegen-II,
-Reid 93, Argonne v$_{18}$ and Paris 1980 (shown in Fig.
\ref{fig3}) are in better agreement with the $y$-scaling data than
those using the charge-dependent Bonn potential (Fig. \ref{fig2}).
This might be related to the fact that these potentials describe
NN phase shifts up to larger energies (e.g. the Nijmegen-II
potential gives reasonable $pp$ phase shifts up to 1.2 GeV, while
the charge-dependent one-boson exchange Bonn potential fits the
phase-shift data below 350 MeV); ii) It can be seen from Fig.
\ref{fig3} that there are small differences between the curves
corresponding to different NN potentials for $k\leq $ 3 fm$^{-1}$
(which give a good description of the $y$-scaling data and almost
coincide with the LFD result) and larger ones for $k>3$ fm$^{-1}$.
Large differences take place, however, between all of them and the
LFD result for $k>3$ fm$^{-1}$.

\begin{figure}[h]
\begin{center}
\includegraphics[width=80mm]{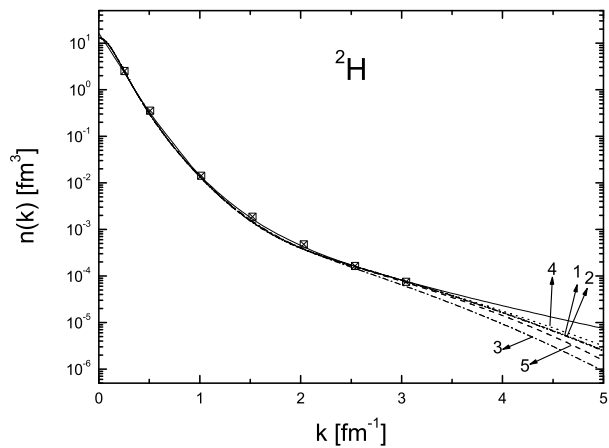}
\vspace{2mm}\caption{The nucleon momentum distribution in deuteron
calculated with the LFD (solid line) in comparison with the
results presented by numerated arrows, as follows: 1-Argonne
v$_{18}$, 2-Nijmegen Reid 93, 3-Nijmegen I, 4-Nijmegen II, 5-Paris
1980 NN potentials and with the $y$-scaling data
\protect\cite{Cio91}.}
\label{fig3}
\end{center}
\end{figure}

\section{Results for the asymptotic scaling function of
deuteron}

In this Section we present the results for the asymptotic scaling
function of deuteron which is calculated by using of the LFD
nucleon momentum distribution given in Section III.

\begin{figure}[h]
\begin{center}
\includegraphics[width=80mm]{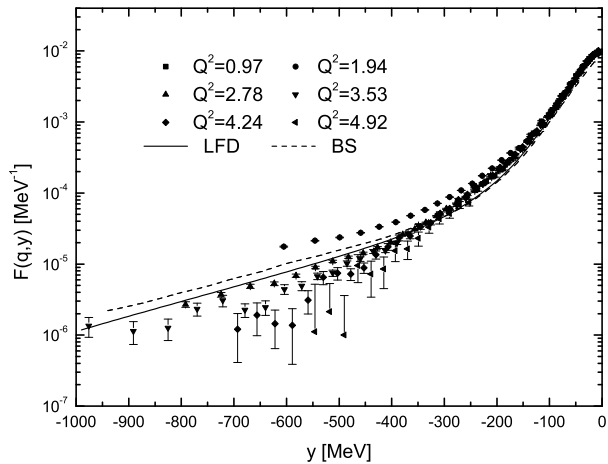}
\vspace{2mm}\caption{The scaling function of deuteron. The
experimental data for different $Q^{2}$ values are from
\protect\cite{Arr2002}. The solid and dashed curves represent the
LFD calculations of this work and BS result of Ref.
\protect\cite{Cio99}.}
\label{fig4}
\end{center}
\end{figure}

The scaling function for deuteron calculated within the LFD method
is shown in Fig. \ref{fig4}. It is compared with the TJNAF
experimental data \cite{Arr2002} for all measured angles. The
$Q^{2}$ values are given for Bjorken $x=Q^{2}/2M\nu=1$ and
correspond to elastic scattering from a free nucleon. It is seen
from Fig. \ref{fig4} that the relativistic LFD scaling function is
in good agreement with the data in the whole region of negative
$y$ available. As known, the scaling breaks down for values of
$y>0$ due to the dominance of other inelastic processes beyond the
quasielastic scattering. Our LFD deuteron scaling function is also
compared in Fig. \ref{fig4} with the scaling function obtained
within the BS formalism \cite{Cio99}. A small difference between
the two results is observed for $y<-400$ MeV/c but, at the same
time, the theoretical LFD scaling function is closer to the
experimental data in the same region of $y$. The fact that both
LFD and BS functions reveal similar behavior is a strong
indication in favor of the consistency of the two relativistic
covariant approaches in case of the $y$-scaling.

\begin{figure}[h]
\begin{center}
\includegraphics[width=80mm]{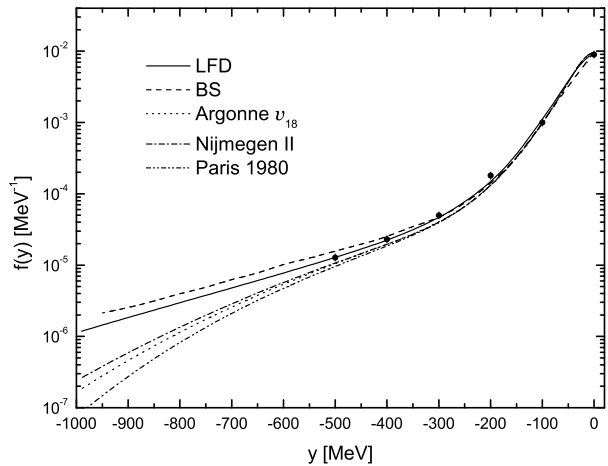}
\vspace{2mm}\caption{The asymptotic scaling function of deuteron.
Line convention refering to calculations using LFD and potential
approaches, as well as BS result \protect\cite{Cio99} is given
(see also the text). The experimental data \protect\cite{Day90}
are given by the full circles.}
\label{fig5}
\end{center}
\end{figure}

In Fig. \ref{fig5} the asymptotic relativistic LFD and BS scaling
functions $f(y)$ are compared with the nonrelativistic ones,
calculated with some realistic interactions. We have already shown
(see Figs. \ref{fig2} and \ref{fig3} of the present work) that the
results using these potentials explain very well (with the
exception of the charge-dependent Bonn potential) all the
available data for $n(k)$ up to $k\simeq 3$ fm$^{-1}$ exactly like
the LFD method. However, it is concluded in \cite{Ant2002} that
for $k>3$ fm$^{-1}$ the LFD results for $n(k)$ deviate strongly
from those of the calculations using NN potentials. Here we would
like to emphasize the existence of the same discrepancies between
the scaling functions observed from the comparison in Fig.
\ref{fig5}. It is shown that for $|y|>$ 400 MeV/c both LFD and BS
curves start to deviate from the nonrelativistic scaling
functions. The result for $f(y)$ calculated using the Nijmegen-II
NN potential is in better agreement with the experimental data
than those using other potentials. It is in accordance with the
result for $n(k)$ shown in Fig. \ref{fig3}. For instance, by a
thorough comparison between the relativistic Bethe-Salpeter and
the nonrelativistic scaling functions of deuteron it has been
found in \cite{Cio99} that the two functions start to sensibly
differ also at $|y|>$ 400 MeV/c. Thus, the necessity to treat
realistically the relativistic dynamics inside the deuteron in a
way different from the potential approaches becomes apparent. In
this sense, the results calculated for both momentum distribution
and asymptotic scaling function confirm the abilities of the LFD
method to describe with a good accuracy the experimental data
measured at high momentum transfers.

\section{Conclusions}

In the present paper inclusive electron-deuteron scattering data
have been analyzed in terms of the $y$-scaling function within the
light-front dynamics method. For this purpose, the nucleon
momentum distribution in deuteron has been used in order to
calculate the asymptotic scaling function. For the trivial case of
deuteron, for which the structure function (Eq. \ref{eq:limit})
coincides with the longitudinal momentum distribution (Eq.
\ref{eq:long}) we have found a good agreement of the calculated
scaling function with the experimental data. Thus, the concept of
relativistic $y$-scaling can be introduced in the LFD relativistic
description of inclusive quasielastic $eD$ scattering, in the same
way as it is done in the conventional nonrelativistic approach,
i.e. by introducing a scaling function (which, in the scaling
regime, is nothing but the nucleon longitudinal momentum
distribution), and in terms of the same variable $y$. It has been
pointed out that for $|y|>$ 400 MeV/c the differences between the
LFD and the nonrelativistic scaling functions are very large.

Exploring the light-front dynamics, we continue in this paper our
analysis of important deuteron characteristics. The effective
inclusion of the relativistic nucleon dynamics and of short-range
NN correlations can be better seen when analyzing electron
scattering at high momentum transfer from complex nuclei, for
which a proper theoretical $y$-scaling analysis is still lacking.
Although scaling violation effects due to final-state interactions
sharply decrease with increasing momentum transfer, a consistent
treatment of both FSI and nucleon binding must be made in order to
perform a precise comparison with the new TJNAF data. Such an
investigation is in progress.

\acknowledgments
This work was partly supported by the Bulgarian
National Science Foundation under the Contracts Nrs.$\Phi $--809
and $\Phi $--905.

\end{document}